\documentclass{optica-article}

\journal{opticajournal} % for journals or Optica Open

\usepackage{siunitx}
\usepackage{graphicx}
\usepackage{braket}
\usepackage{mathrsfs}
\usepackage{amsmath}

\articletype{Research Article}
\usepackage{lineno}
\begin{document}

\title{Time-Bin Properties of Photon Pairs from a 2GHz Waveguide Resonator}

\author{Stefan Kazmaier\authormark{*} and Kaisa Laiho}

\address{German Aerospace Center (DLR e.V.), Institute of Quantum Technologies, Wilhelm-Runge-Str. 10, 89081 Ulm, Germany}
\email{\authormark{*}stefan.kazmaier@dlr.de} %% email address is required; see note below about the corresponding author designation
% use {asbstract*} to suppress the copyright line. Copyright information will be added in production

\begin{abstract*} 
We produce degenerate photon pairs called signal and idler via parametric down-conversion (PDC) in a Fabry-Pérot waveguide resonator having close to a \SI{2}{GHz} bandwidth in the telecom C-band. Our PDC source is based on a type-II periodically-poled Lithium Niobate waveguide with high-reflecting end facets forming the resonator. We measure the strength of the photon-pair correlation between signal and idler after selecting a single resonator mode from the multimode PDC emission. Thereafter, we investigate their close to one nanosecond long pulse shapes, in other words, the time bins, with standard photon counting.
We find that the time-bin durations vary significantly from each other, which can be predicted by the different optical losses of the cross-polarized signal and idler. Further, we simulate the temporal properties of signal and idler via the Fourier transform of their joint spectral characteristics. Our results deliver empiric insight about the time bins of photon pairs and are relevant for driving future quantum networks at the edge to narrow bandwidths but nevertheless in the pulsed regime.
\end{abstract*}

\section{Introduction}

The future hybrid quantum networks incorporating both terrestrial and aerospace segments require versatile and highly tuneable optical hardware components. The requirements of such network components may embody the telecommunication wavelength range in order to simultaneously hit the low-loss transmission window of optical fibers and to adequately send light through the atmosphere \cite{saleh1991, Liao2017, Khatri_2021}. Another stringent condition is delivered by the trade-off between the frequency and time domains of optical pulses, which are connected via the Fourier transform \cite{siegman1986}. Pulsed photonic sources are interesting, since they offer intrinsic means for the synchronization of large networks \cite{Quan.2016}. Additionally, pulsed applications would allow for a fast operation, since the transmission rate is limited by the time-bin duration. However, due to their broad spectral bandwidths, ultrafast photonic sources are hardly compatible with other network components, such as quantum memories that often rely on spectrally sharp elementary transitions. Regarding the telecommunication wavelength range, recent developments in quantum memories have shown a GHz-scale bandwidth on a rear-earth-ion-doped platform incorporating Erbium \cite{Rancic2018, Barya2026}. In this spectral regime also  a memory protocol based on acoustic waves has been demonstrated  at the few photon level \cite{Saffer_2025}. \\
We investigate a Fabry-Pérot (FP) waveguide (WG) resonator made of periodically-poled Lithium Niobate (PPLN) as a source of photon pairs in order to approach the region of narrower bandwidths but remain in pulsed operation. The optical losses inside our WG resonator limit its bandwidth, which is close to \SI{2}{GHz} in the full width at half maximum (FWHM) \cite{Kazmaier_2025}. In the past FP PPLN WG resonators have been shown to emit quantum light into a single optical mode and to be suitable for the heralded single-photon generation \cite{Luo_2015} as well as for the generation of time-energy entanglement \cite{Pomarico2009}. In contrast to their bulk counterparts, waveguided resonator sources offer a higher brightness, a better potential for further optical integration, and typically these resources are passive without the need for active stabilization of the resonator \cite{Slattery2019, Mataji-Kojouri2023}. \\ %Historically, Fabry-Perot waveduide resonators  are excellent sources for also lasing \cite{Lin2004,Zhao2016, Fang2010}. ...
The spectro-temporal properties of the parametric down-conversion (PDC) stem from the energy and momentum conservation of the underlying second-order non-linear optical process \cite{Brecht2013}. Typically, the PDC sources that are pumped with pulsed lasers result in several nanometers broad bandwidths in the spectral domain. In such a case, their joint-spectral properties can be measured with standard optical equipment \cite{Avenhaus2009}. However, the duration of the signal and idler time bins then ends up well below the detector's jitter level that typically lies around $\SI{20}{ps}$ for state-of-the-art single-photon counters. In such cases, only indirect measurements of the temporal domain can reveal the pulse shapes of signal and idler \cite{Koviri2025, PhysRevLett.101.153602, Allgaier_2017}. Nevertheless, a careful engineering of the temporal shape of photons is crucial in quantum optics. Usually, this is required for overlapping independent but indistinguishable pulses \cite{Laiho2009}, whereas tasks such as achieving an optimized atom-photon coupling lie at the high-end \cite{Golla2012}.\\
Our FP PPLN WG resonator is realized by coating the WG's end-facets with a high reflection coating. This photonic source is designed to emit degenerate photon pairs around \SI{1538}{nm}. Our main objective is to study the temporal pulse shapes of signal and idler that we call the time bins. 
In case signal and idler are emitted in multiple resonator modes (MRMs) we further  
gain information of the resonator characteristics themselves \cite{PhysRevLett.123.193603, Murakami2025}. 
This paper is organized as follows. First, a theoretical treatment of the generated PDC emission in our FP PPLN WG resonator is given. Second, we measure the coincidences-to-accidentals ratio (CAR) to prove the photon-pair creation within a single resonator mode (SRM), which is realized by spectral filtering. Third, we measure the time bins of signal and idler both in MRM and SRM cases and see distinct differences between them. Moreover, we observe very different time-bin durations for the cross-polarized signal and idler, which can be predicted by polarization dependent optical losses inside the waveguide.
Our results show a detailed study of our PDC source in the temporal domain and highlight the influence of the underlying process parameters on the temporal degree of freedom of the PDC emission.
%%%%%%%%%%%%%%%%%%%%%%%%%%%%%%%%%%%%%%%%%%%%%%%%%%%
\section{Theoretical treatment}
For the simulation of the temporal properties of the photon pairs produced in the PDC a thorough theoretical treatment of the joint spectral intensity (JSI) is needed. Generally, a photon-pair state $\ket{\psi}$ generated via PDC can be defined as \cite{PhysRevA.56.1627, PhysRevA.56.1534} 
\begin{equation}
    \ket{\psi} \approx \ket{0} + \epsilon \int d\omega_\mathrm{s} \int d\omega_\mathrm{i} \gamma(\omega_\mathrm{s}+\omega_\mathrm{i}) \mathrm{sinc}\left(\frac{L}{2}\Delta k(\omega_\mathrm{s},\omega_\mathrm{i})\right) \exp{\left( i\frac{L}{2}\Delta k(\omega_\mathrm{s},\omega_\mathrm{i})\right)}\ket{\omega_\mathrm{s}}\ket{\omega_\mathrm{i}}\label{eq:psi}
\end{equation}
with $|\epsilon|^2\ll1$ being related to the probability of preparing a photon-pair state. In Eq.~(\ref{eq:psi}) the parameter $L$ denotes the interaction length, which is equal to the PPLN WG length here. The properties of the signal-idler wavepacket stem from the interaction of the pump ($\mathrm{p}$), signal ($\mathrm{s}$) and idler ($\mathrm{i}$) beams. They can be expressed in terms of the phasematching function, pump spectral amplitude, and resonator response function. Signal and idler are expressed in terms of the frequency-states $\ket{\omega_\mathrm{j}}=\hat{a}^\dagger_\mathrm{j}(\omega_\mathrm{j})\ket{0}$ ($\mathrm{j} = \mathrm{s},\mathrm{i}$), in which $\hat{a}^{\dagger}_\mathrm{j}(\omega_\mathrm{j})$ is the photon creation operator of the propagating mode at the angular frequency $\omega_\mathrm{j} =2\pi c/\lambda_\mathrm{j}$, with $\lambda_\mathrm{j}$ denoting the wavelength, and $c$ representing the speed of light.
Due to energy conservation $\omega_\mathrm{p} = \omega_\mathrm{s} +\omega_\mathrm{i}$.
In Eq.~(\ref{eq:psi}) the phasematching condition is denoted by $\Delta k (\omega_\mathrm{s},\omega_\mathrm{i}) = k_\mathrm{p}(\omega_\mathrm{p} =\omega_\mathrm{s} +\omega_\mathrm{i})-k_\mathrm{s}(\omega_\mathrm{s})-k_\mathrm{i}(\omega_\mathrm{i})-2\pi/\Lambda$, in which $k_{\upmu} = \frac{n_{\upmu}(\omega_\mathrm{\upmu})\omega_{\upmu}}{c}$ ($\upmu = \mathrm{s}, \mathrm{i}, \mathrm{p}$) is given in terms of the dispersion relation of the interacting modes and $\Lambda$ is the poling period. Furthermore,  the pump spectral amplitude $\beta$ and the resonator response function $A$ are summarized in $\gamma$ in Eq.~(\ref{eq:psi}) via
\begin{align}
    \gamma(\omega_\mathrm{s}+\omega_\mathrm{i})&=\beta(\omega_\mathrm{s}+\omega_\mathrm{i}) A_\mathrm{s}(\omega_\mathrm{s}) A_\mathrm{i}(\omega_\mathrm{i}), \ \textrm{in which}\\
    \beta(\omega_\mathrm{s}+\omega_\mathrm{i})&=\exp{\left(-\frac{1}{\sigma_\mathrm{p}^2}(\omega_\mathrm{s}+\omega_\mathrm{i}-\omega_\mathrm{p})^2\right)}\ \label{eq:pump} \textrm{and}\\
    A_\mathrm{j}(\omega_\mathrm{j}) &= \frac{\sqrt{(1 - R_\mathrm{j,1})(1 - R_\mathrm{j,2})} e^{-\alpha_\mathrm{j} L / 2}}{1 - \sqrt{R_\mathrm{j,1} R_\mathrm{j,2}} e^{-\alpha_\mathrm{j} L} e^{i \phi_\mathrm{j}(\omega_\mathrm{j})}} \label{eq:resonator}.
\end{align}
In Eq.~(\ref{eq:pump}) the pump spectrum $\beta$ is approximated with a Gaussian function with its bandwidth related to $\sigma_\mathrm{p}$. 
Following Ref.~\cite{Brecht2016}, the FP Resonator is treated classically, leading to the response function in Eq.~(\ref{eq:resonator}) with the roundtrip phase $\phi_\mathrm{j}(\omega_\mathrm{j})=2\omega_\mathrm{j} n_{\mathrm{j}}(\omega_\mathrm{j}) L/c$. 
The transmission spectrum of the resonator is proportional to $|A(\omega)|^2$. 
In the following a two-dimensional Fourier transformation will be used to transform from the frequency domain into the time domain. This yields
\begin{equation}
    \mathscr{F}(t_\mathrm{s},t_\mathrm{i})
    = \iint f(\omega_\mathrm{s},\omega_\mathrm{i})
    \exp\left[-i2\pi
    \left(t_\mathrm{s}\omega_\mathrm{s}+t_\mathrm{i}\omega_\mathrm{i}\right)\right]
    \,d\omega_\mathrm{s}\,d\omega_\mathrm{i},
\end{equation}
 which is given in terms of the doublet ($t_\mathrm{s}, t_\mathrm{i}$) that denotes the emission times for signal and idler, respectively.
\begin{table}[tp]
\caption{Parameters for simulating the joint spectral properties partly stemming from experiments reported in \cite{Kazmaier_2025}. The idler beam is polarized along the extraordinary ($\mathrm{e}$) crystal axis, while the signal and pump beams are polarized along the ordinary ($\mathrm{o}$) axis.}
  \centering
\begin{tabular}{cc}
\hline
Parameter & Value  \\
\hline
Pump pulse length & \SI{0.4}{ps}  \\
Poling period & \SI{9}{\micro m} \\
PPLN temperature & \SI{50}{\degree C} \\
PPLN length & \SI{10.30\pm0.05}{mm}\\
Resonator reflectivity & $R_1=\SI{97+-2}{\%}; R_2=\SI{97.7+-0.5}{\%}$\\
Bandwidth (meas.) & $\delta_\mathrm{e}=$\SI{2.29\pm0.01}{GHz}; $\delta_\mathrm{o}=$\SI{2.14\pm0.01}{GHz} \\
FSR (meas.) & $\nu^\mathrm{FSR}_\mathrm{e}=$\SI{6.70\pm0.01}{GHz}; $\nu^\mathrm{FSR}_\mathrm{o}=$\SI{6.45\pm0.01}{GHz} \\
\hline
\end{tabular}
\label{tab:SimulatedParameters}
\end{table} 
The simulation parameters are summarized in Tab.~\ref{tab:SimulatedParameters}.
The resonator bandwidth and its decay time are strongly influenced by the linear optical loss $\alpha_\mathrm{j}$. We search for a qualitative agreement between the theory and experiment in order to predict the signal and idler time bin properties. In the simulation $\alpha_\mathrm{j}$ is chosen such that the theory matches with the experiment. Therefore, the parameters in the SRM and MRM cases slightly vary from each other. 
We use the linear loss parameters $\alpha_\mathrm{s}^\mathrm{SRM}=\SI{0.27}{dB/cm}$ and $\alpha_\mathrm{i}^\mathrm{SRM}=\SI{2.5}{dB/cm}$ as well as $\alpha_\mathrm{s}^\mathrm{MRM}=\SI{0.3}{dB/cm}$ and $\alpha_\mathrm{i}^\mathrm{MRM}=\SI{2.0}{dB/cm}$.  

For calculating the phasematching function we use Sellmeier's equation \cite{Sellmeier1872} with the parameters found in Ref.~\cite{Jundt:97} for predicting the refractive indices. A small correction to these values is needed to match the resonator's free-spectral range (FSR) between simulation and experiment. A direct connection between the dispersion relations and the FSR exists via $\nu^\mathrm{FSR}_\mathrm{j}\approx\frac{c}{2n_\mathrm{g,j}L}$ in which $n_\mathrm{g,j}$ is the group refractive index.
Therefore, the FSR can be used to adapt the refractive index model to the experiment. Here, the simulated refractive index is reduced approximately by 4\textperthousand \ for the extraordinary and ordinary crystal axis, leading to a correction of 3-4\textperthousand \ in the FSR. The properties of the guided modes strongly depend on the underlying waveguide design and the fabrication tolerances are known to influence the dispersion of the integrated optics devices.
% In waveguided structures the interaction of a guided mode with the surrounding material and fabrication tolerances are known to influence the dispersion of guided modes in the integrated optics devices. 
Therefore, often a correction to the calculated refractive index properties is necessary \cite{PhysRevA.80.033829, Iveson:26}.
%\textcolor{red}{Should we put in a relative or absolute correction? Is a correction at the pump wavelength necessary?} 
Our earlier experiment also indicates the necessity of a small correction to the values predicted by the Sellmeier's equations \cite{Kazmaier_2025}.

\begin{figure}[tbp]
\centering\includegraphics[width=\linewidth]{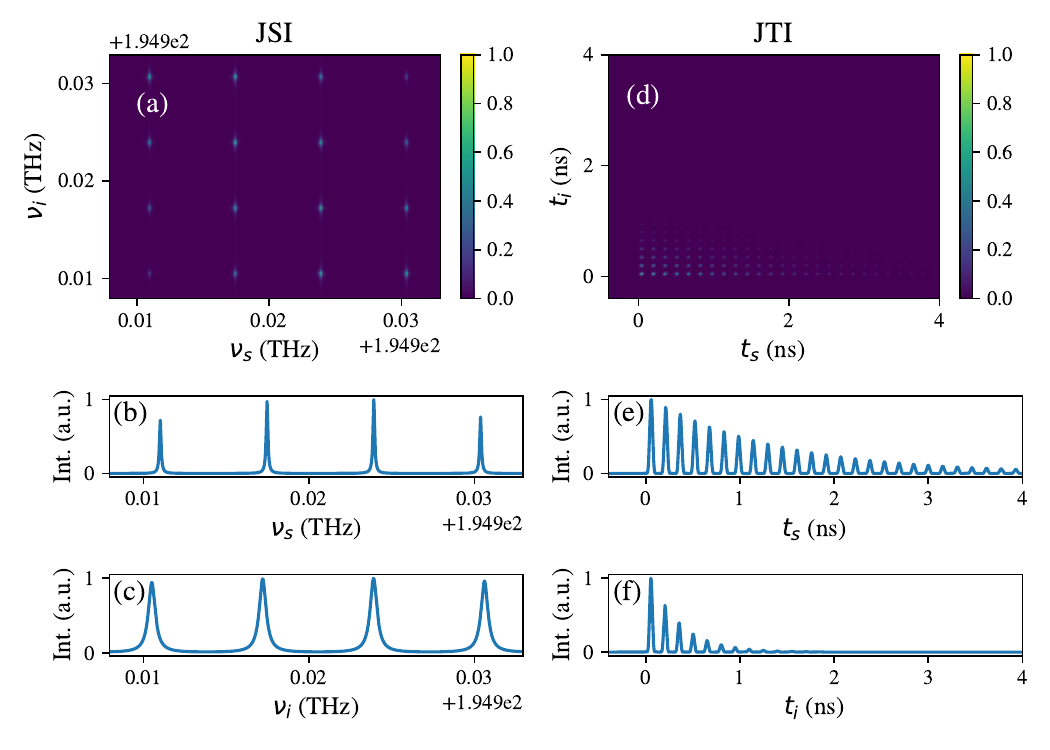}
\caption{Simulation in the MRM case. (a) presents the JSI in terms of the signal and idler frequencies $\nu_\mathrm{s,i}$ while (b) \& (c) show the spectral marginal distributions of signal and idler, respectively. (d) illustrates the 2D Fourier transform into the time domain resulting in the JTI, while (e)  \&  (f) show the time bins of signal and idler, respectively. We have limited the frequency scales in (a) and the time scale in (b) for better visualization of the resonator modes.}
\label{fig: SMC_Multimode_JSI_FFT_with_maringals}
\end{figure}
For the MRM case we present in Fig.~\ref{fig: SMC_Multimode_JSI_FFT_with_maringals}(a) the joint spectral intensity (JSI) given by the product in Eq.~(\ref{eq:psi}) in terms of the signal and idler frequencies $\nu_\mathrm{s,i}$. 
The spectral marginal distributions of signal and idler are presented in Fig.~\ref{fig: SMC_Multimode_JSI_FFT_with_maringals}(b,c). The different envelopes in the marginal distributions of the JSI stem from the fact that the phase matching function is not perfectly anticorrelated such as the pump spectrum is.
By performing the 2D Fourier transform we gain the joint temporal intensity (JTI) of signal and idler that is shown in Fig.~\ref{fig: SMC_Multimode_JSI_FFT_with_maringals}(d). 
% By evaluating the temporal marginal distributions, in other words, the time bins of signal and idler in Figs. \ref{fig: SMC_Multimode_JSI_FFT_with_maringals}(e,f) we notice that in the time domain the pulse envelope takes the form of an exponential decay with a fringe pattern due to the interference of the multiple spectral modes.
The temporal marginal distributions, in other words, the time bins of signal and idler are shown in Fig.~\ref{fig: SMC_Multimode_JSI_FFT_with_maringals}(e,f). We notice that in the time domain the pulse envelope takes the form of an exponential decay with a fringe pattern due to the interference of the multiple spectral modes. 
We note that the different decay times of the signal and idler modes are predicted just by the different optical losses. 
\begin{figure}[tbp]
\centering\includegraphics[width=\linewidth]{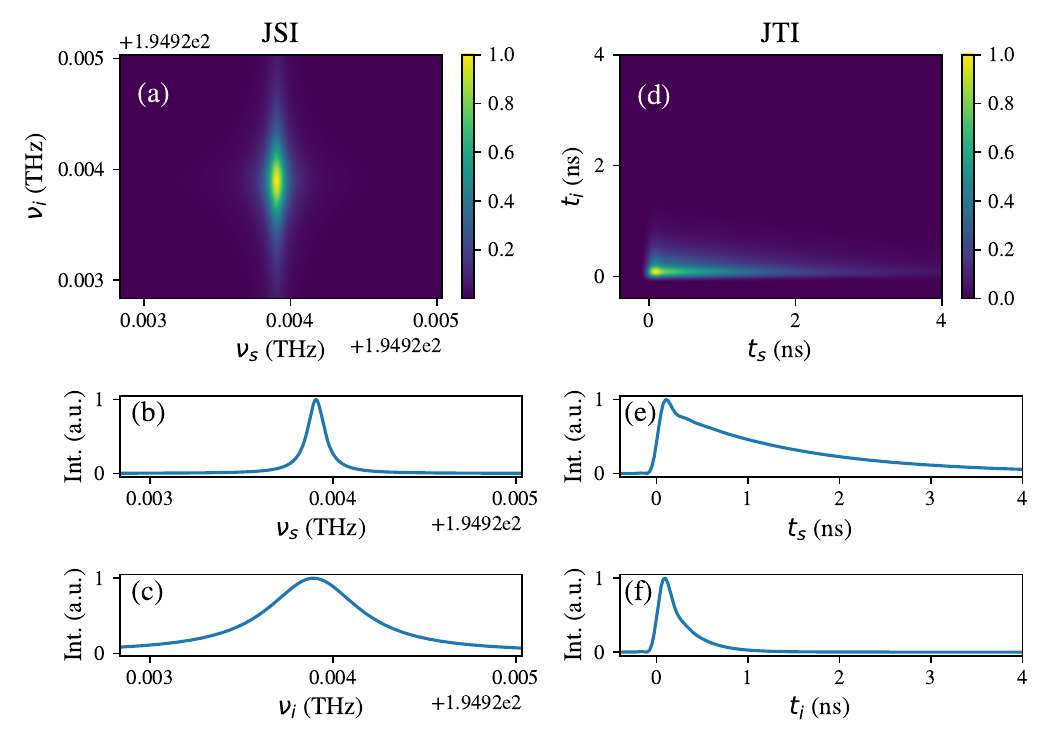}
\caption{Simulation in the SRM case. (a) presents the JSI in terms of the signal and idler frequencies $\nu_\mathrm{s,i}$ while (b)  \&  (c) show the spectral marginal distributions of signal and idler, respectively. (d) illustrates the 2D Fourier transform into the time domain resulting in the JTI, while (e)  \&  (f) show the time bins of signal and idler, respectively. }
\label{fig: SMC_Singlemode_JSI_FFT_w_Marginals}
\end{figure}

Thereafter, the SRM case is simulated by choosing a small frequency window around a single spectral resonator peak, as similarly shown in the spectral domain in Fig.~\ref{fig: SMC_Singlemode_JSI_FFT_w_Marginals}(a-c). Finally, its 2D Fourier transform to the time domain with the temporal marginal distributions are shown in Fig.~\ref{fig: SMC_Singlemode_JSI_FFT_w_Marginals}(d-f). While the exponential decay remains in the time bins, the fringe pattern has disappeared.\\
For the calculations a grid of $5\cdot10^3 \times  5\cdot10^3$ entries is used to simulate in the frequency domain of the joint spectral properties. Before the Fourier transform, zero padding is applied, through which the grid size is increased to $2\cdot10^4 \times  2\cdot10^4$. Additionally, a Tukey window with shape parameter 0.8 is used to minimize sinc contributions in the phasematching function. It is noted that the use of a window function affects the shape of the temporal decay, since it acts as an additional envelope for the peak spectrum. Thus, it has to be chosen carefully.
With this approach of applying two well-established numerical methods we aim at reducing numerical artifacts in the Fourier transformation \cite{Thiel_2023, Kazmaier_2025}.
%%%%%%%%%%%%%%%%%%%%%%%%%%%%%%%%%%%%%%%%%%%%%%%%%%%
\section{Experimental arrangement}
\begin{figure}[tbp]
\centering\includegraphics[width=0.9\linewidth]{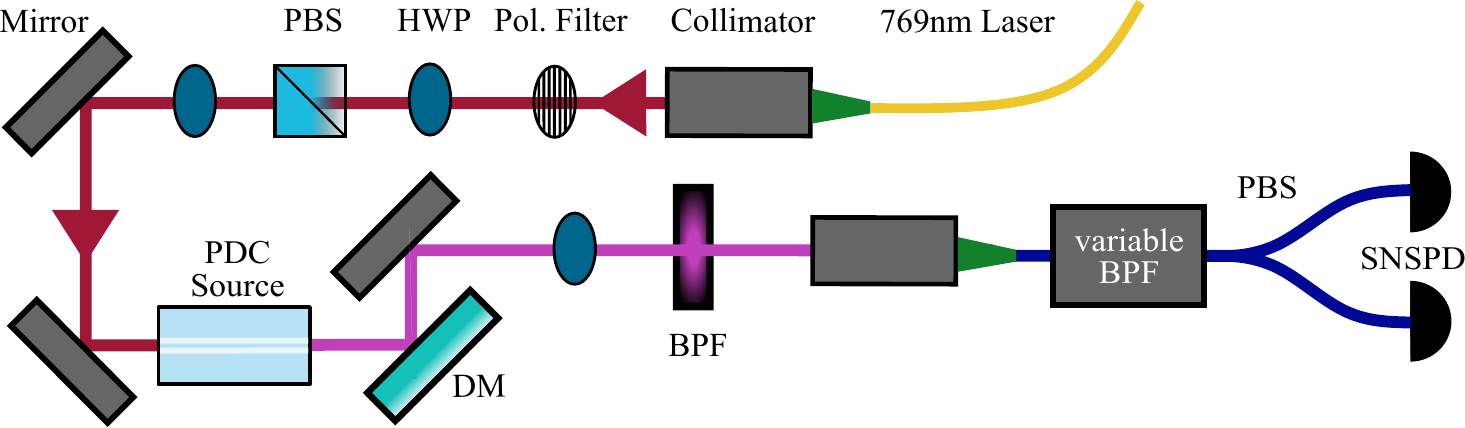}
\caption{Experimental arrangement. For more details see the main text. Abbreviations:  Pol. filter = polarization filter; HWP = half-wave plate; PBS = polarizing beam splitter; DM = dichroic mirror; BPF = bandpass filter; SNSPD = superconducting nanowire single photon detector.}
\label{fig: Experimental Setup}
\end{figure}

Our experimental arrangement for measuring the signal and idler characteristics is shown in Fig.~\ref{fig: Experimental Setup}.
The pump light at \SI{769}{nm} is produced by a pulsed laser system (Stuttgart Instruments) running at the repetition rate of \SI{40}{MHz} with a pulse duration of \SI{0.4}{ps}.
The pump beam passes through an optical fiber, a polarization filter, and a half-waveplate (HWP) in front of a polarizing beam splitter (PBS) to control its spatial mode, polarization, and power, respectively. Thereafter, another HWP adjusts the polarization to match with the one required for the PDC process.
The pump beam is coupled into and out of the PPLN WG resonator with aspheric lenses having the focal lengths of \SI{9.6}{mm} and \SI{7.5}{mm}, respectively.
The WG structure on the PPLN platform is realized with air gaps to the left and right of the core region and reflecting oxide layers on top and bottom to allow only for the fundamental transverse mode to propagate. This gives an asymmetric layout with respect to the polarization directions of the cross polarized signal and idler. The end-facets of the PPLN WG are coated with a high-reflection coating in order to form a FP resonator at the PDC emission wavelengths close to \SI{1538}{nm}. An anti-reflection coating is applied around \SI{769}{nm} to achieve single-pass operation at the pump wavelength. A more detailed description of the PPLN WG resonator is reported in Ref.~\cite{Kazmaier_2025}.\\
Therafter the pump beam is removed from the PDC emission by using a dichroic mirror with a cut-off edge close to \SI{1180}{nm} and a free-space bandpass filter centered at \SI{1540}{nm} having a bandwidth of \SI{12}{nm}. Another HWP is used to control the signal and idler polarization before being coupled into a polarization maintaining telecom fiber. 
Optionally, a fiber coupled polarization maintaining variable bandpass filter with wavelength resolution of \SI{5}{pm} and minimum bandwidth of \SI{32}{pm} is connected. This filter reduces the number of spectral modes in the generated PDC emission. It can also filter out only one of the resonator modes. 
Afterwards, the photons are send through a polarization controller (not shown in Fig.~\ref{fig: Experimental Setup}) and split into signal and idler using a fiber integrated polarizing beam splitter. The recorded counts are maximized by optimizing the polarization of light (not shown in Fig.~\ref{fig: Experimental Setup}) and detected with superconducting nanowire single-photon detectors (SNSPDs). Signal and idler arrival times are obtained using a time tagger, which is triggered by the pulsed pump laser, a small portion of which is detected with a fast photodiode (not shown in Fig.~\ref{fig: Experimental Setup}). In the time-bin measurements, we expect a total timing jitter below \SI{80}{ps} in FWHM. This jitter has been measured separately with faint pulsed light from our laser system in the telecommunication wavelength range, and it incorporates the jitters of the fast photodiode used as a trigger, the used SNSPD channel and two used time-tagger channels.
% for each used SNSPDs channel,which incorporates the jitters of the fast photodiode used as a trigger (<\SI{22}{ps}), that of the used SNSPD channel (<\SI{70}{ps}) and that of the two time-tagger channels (<\SI{17}{ps}) used. 
We expect to measure the convolution between the detection system response and the temporal pulse envelopes. 
However, later on we see that the convolution effect in our results is minor.
% However the resolution of the fringes with a spacing of \SI{16}{ps} in the MRM case show that the convolution effect is negligible. 
Nevertheless, the jitter defines the resolution limit and the accuracy of our time-bin measurements.
%%%%%%%%%%%%%%%%%%%%%%%%%%%%%%%%%%%%%%%%%%%%%%%%%%%
\section{Results and discussion}
\begin{figure}[tbp]
\centering\includegraphics[width=0.6\linewidth]{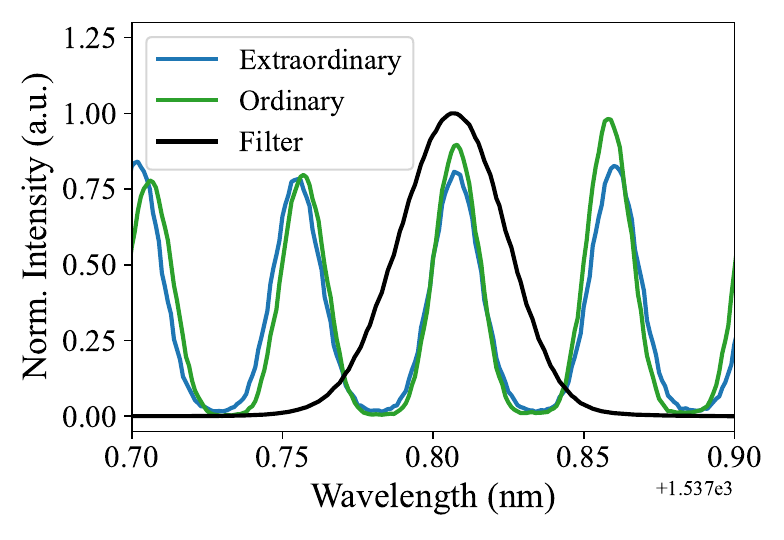}
\caption{Linear transmission measurement for the extraordinary and ordinary crystal axis of the PPLN WG resonator together with the transmission window of the fiber coupled variable bandpass filter. The latter is applied in order to filter a SRM from the PDC emission.}
\label{fig: Cluster and Filter}
\end{figure}

Prior to measuring the PDC emission we analyze the bandwidth of the PPLN WG resonator and set the transmission window of the fiber-based variable filter accordingly. For this purpose we send a broadband, pulsed telecom laser through the PPLN WG resonator and detect the transmitted light with an optical spectrum analyzer similar to the measurements reported in Ref.~\cite{Kazmaier_2025}. The transmission of light polarized along both the extraordinary and ordinary crystal axes is measured around \SI{1538}{nm} and the typical resonator pattern is recorded. Since the group indices for these two axes are different, their free spectral ranges (FSRs) slightly vary from each other and only a few peaks overlap forming a cluster. Our results are shown in Fig.~\ref{fig: Cluster and Filter} together with the selected transmission window for the variable filter.
\subsection{Determining the strength of photon-pair correlation}
\begin{figure}[tbp]
\centering\includegraphics[width=0.8\linewidth]{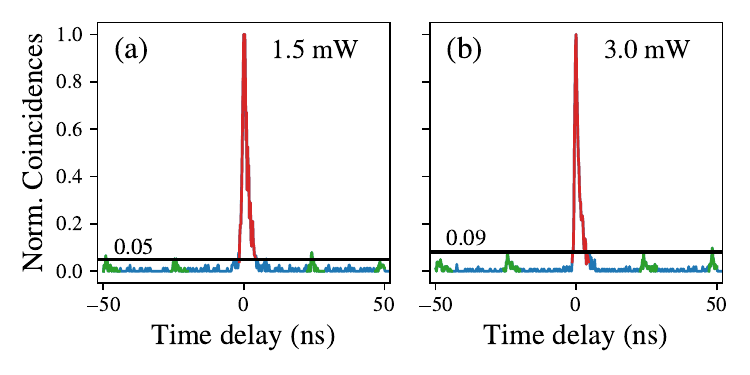}
\caption{Histogram of coincidence counts occurring between signal and idler at the pump powers of  (a) \SI{1.5}{mW} and (b) \SI{3.0}{mW}. The counts in the central peak (red) contribute to the coincidences, while the side peaks (green) contribute to the accidentals with a black horizontal line as a guide to the eye. The spurious counts that lie outside the considered time gates are colored in blue.
}
\label{fig: Histogram for CAR}
\end{figure}

We start by measuring the strength of the photon-pair correlation by determining the well-known CAR. For that purpose we discriminate the coincidence counts between signal and idler in the SRM case and make a temporal start-stop measurement between signal and idler at two different pump powers of \SI{1.5}{mW} and \SI{3.0}{mW}. Our results in Fig.~\ref{fig: Histogram for CAR} show a central peak of coincidence counts, if photons are produced in pairs. The side peaks arrive from the accidental counts that match with photons created by the subsequent pump pulses. Due to the used spectral filtering spurious counts from the resonator side peaks can also contribute to the accidentals.
We extract the values of $\mathrm{CAR}_\mathrm{1.5mW}=\SI{31.3\pm1}{}$ and $\mathrm{CAR}_\mathrm{3mW}=\SI{16.8\pm0.2}{}$. In the evaluation we use \SI{6}{ns} long time gates in order to reduce the effect of background counts.
As expected for the PDC process, the CAR value roughly halves, when the  pump power is doubled \cite{Chen2018}. The values for the CAR reached in our case are comparable with results reported in the past for platforms generating PDC emission in guided-mode resonators \cite{Huang_2025_10.1063/5.0263312,Luo_2015, Guo2017}. We further emphasize that it is important to reach and remain in the region of values CAR\SI{\gg 2}{}, such that a stronger photon-pair correlation than in thermal light can be proven.
\subsection{Investigation of MRM time bins}
\begin{figure}[tbp]
\centering\includegraphics[width=0.9\linewidth]{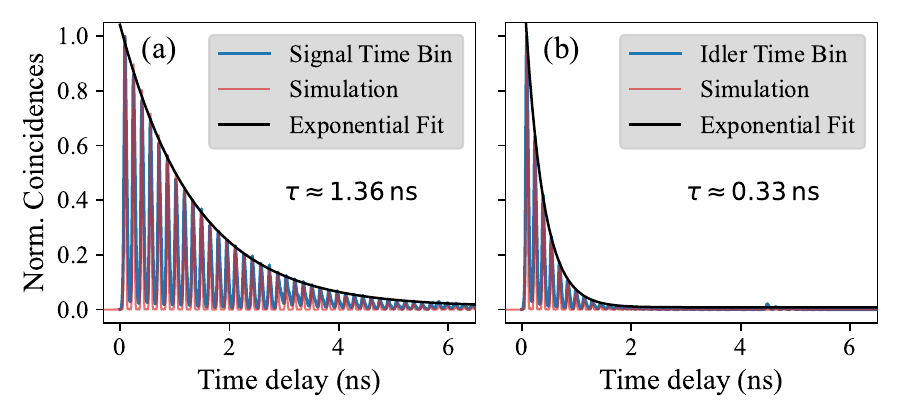}
\caption{Measured time bins (blue lines) of (a) signal and (b) idler in the MRM case together with the extracted decay times $\tau$. The exponential decays (black lines) are fitted to the envelope of the peak pattern. The simulated temporal behavior (red lines) is gained from a 2D Fourier transform of the joint spectral properties. The fringe pattern is a clear sign of the multimode operation.}
\label{fig: Decay Multimode}
\end{figure}

Next, we measure the time bins of signal and idler in the MRM case. For that purpose we remove the fiber-based variable filter from the setup in Fig.~\ref{fig: Experimental Setup}. We utilize the signal from the laser's fast photodiode as start and an event in either signal or idler as stop in the time tagging. In other words, we measure the temporal distribution of the singles' counts.
Our results are shown in Fig.~\ref{fig: Decay Multimode}. An exponential decay proportional to $\exp(-t/\tau)$ is used as a fit for the measured peaks, in which $t$ is the delay time and $\tau$ is the decay time. The decay time corresponds to the temporal duration in which the signal amplitude has fallen to 1/e times its maximum. Our results deliver 
% $\tau_\mathrm{s}= \SI{1.36\pm0.04}{ns}$ and $\tau_\mathrm{i}= \SI{0.33\pm0.04}{ns}$ 
$\tau_\mathrm{s}\approx \SI{1.36}{ns}$ and $\tau_\mathrm{i} \approx \SI{0.33}{ns}$ 
%$\tau_\mathrm{s}\approx \SI{0.94+-0.08}{ns}$ and $\tau_\mathrm{i} \approx \SI{0.23+-0.08}{ns}$
for signal and idler, respectively. We notice that the error from the fit to the experimental data \SI{+-16}{ps} and \SI{+-5}{ps} is much lower than the total jitter in the measurements. Therefore, we expect that the accuracy of these values is determined rather by the total jitter than by the fitting error.
The fringe pattern occurs due to an interference of the multiple spectral resonator modes in the PDC emission. By employing the 1D Fourier transform from the time domain into the frequency domain the FSR of the resonator can be determined to be $\mathrm{FSR} \approx \SI{6.7\pm1}{GHz}$, which shows an excellent agreement with our earlier linear transmission measurements \cite{Kazmaier_2025}. 
The interference pattern delivered by the 2D Fourier transformation of the simulated joint spectral properties in the MRM case shows excellent agreement with the measurement. The slight discrepancy is expected to result from the timing jitter of the detection system, which is not included in the simulation.
\subsection{Investigation of SRM time bins}
\begin{figure}[tbp]
\centering\includegraphics[width=0.9\linewidth]{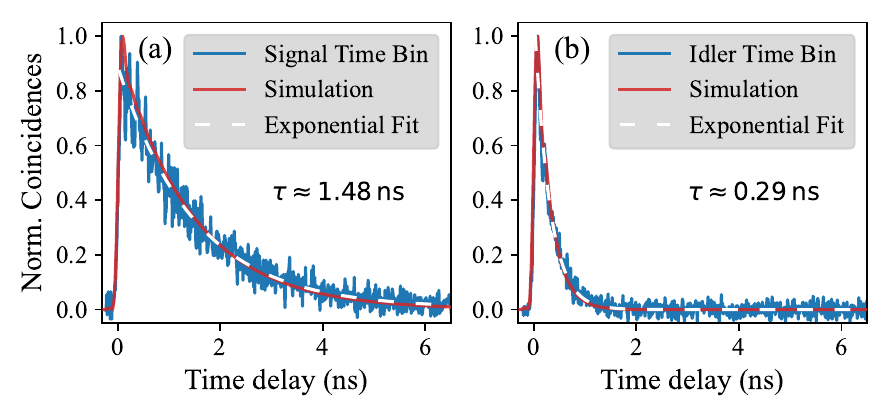}
\caption{Measured time bins (blue lines) of (a) signal and (b) idler in the SRM case together with the extracted decay times $\tau$. Exponential decays (white dashed lines) are fitted to the measurement data. The simulated temporal behavior (red lines) is gained from a 2D Fourier transform of the joint spectral properties.
}
\label{fig: Decay Singlemode}
\end{figure}
Finally, we measure the temporal marginal distributions of signal and idler in the SRM case. For that purpose, we use the setup presented in Fig.~\ref{fig: Experimental Setup} with the spectral filtering shown in Fig.~\ref{fig: Cluster and Filter}. Our results are shown in Fig.~\ref{fig: Decay Singlemode}. First of all, we notice that the fringe pattern disappears, since there no longer is interference between multiple spectral modes, but the PDC is emitted to one individual resonator mode only. 
Our results deliver the decay times 
% (change only slightly ... which could be explained by the fit sitting between the measured values instead of only ontop of the peaks / remain roughly equal) to the MRM case with 
% $\tau_\mathrm{s} = \SI{1.48\pm0.04}{ns}$ and $\tau_\mathrm{i} = \SI{0.29\pm0.04}{ns}$ 
$\tau_\mathrm{s}\approx \SI{1.48}{ns}$ and $\tau_\mathrm{i} \approx \SI{0.29}{ns}$ 
for signal and idler, respectively. We notice that the error from the fit to the experimental data \SI{+-27}{ps} and \SI{+-5}{ps} is again lower than the total jitter in the measurements.
In the SRM case we also predict the temporal pulse shapes of signal and idler by performing a 2D Fourier transformation of the simulated joint spectral properties. Again, we find that the simulated temporal shapes, in which we use the different optical losses for the cross-polarized signal and idler beams, agree very well with the measured time bins. We emphasize that the temporal shapes predicted by the simulation only require adapting the loss parameter. \\
Finally, in Tab.~\ref{tab:Comparison} we compare our results with some experiments reporting the temporal properties of the PDC emission on selected platforms. While ultrafast systems can be probed by up-conversion measurements \cite{Allgaier_2017}, nanosecond systems are close to continuous-wave operation \cite{10.1063/5.0263312, Luo_2015, PhysRevLett.123.193603, Guo2017}. Our system is in the pulsed regime, and a direct measurement with the state-of-the-art photodetection at the single-photon level is possible.

\begin{table}[tbp]
\caption{Summary of some comparable experiments on similar and different platforms and resonator types. *calculated from decay time, $^+$estimated from a data plot, $^\mathsection$coherence time, $^\dag$transmission measurement, **pulse duration extracted from up-conversion measurements, WGMR: Whispering Gallery Mode Resonator, MRR: Microring Resonator, AlN: Aluminum Nitride, KTP: Potassium Titanyl Phosphate}
\centering
\begin{tabular}{c|c|c|c|c}
\hline
Reference & Resonator-type  & Wavelength  & Bandwidth & Decay time \\ 
& \& Platform &(nm)&(GHz)&(ns) \\
\hline
Huang 25\cite{10.1063/5.0263312} & WGMR - LN & 950 \& 1209 & (2.4 \& 3.4)$\cdot$\SI{e-3}{} * & 66 \& 47 \\
Luo 15 \cite{Luo_2015} & FPR - PPLN & 890 \& 1320 & 0.066 * & 3 \& 2.4  \\
Ikuta 19 \cite{PhysRevLett.123.193603} & FPR - PPLN & 1600 & 0.114 * & $\approx$ 2.0 - 2.2 $^+$ \\
Guo 17 \cite{Guo2017} & MRR - AlN & 1550 & \SI{1.1}{} * & 0.145 $^\mathsection$ \\
This work & FPR - PPLN & 1538 & \SI{2.14}{} \& \SI{2.29}{} $^\dag$ & 1.36 \& 0.33  \\
Allgaier 17 \cite{Allgaier_2017} & none - PPKTP & 1545 \& 1543 &  \SI{966}{} & \SI{1.1}{}$\cdot$\SI{e-3}{} ** \\ 
\hline
\end{tabular}
\label{tab:Comparison}
\end{table}
%%%%%%%%%%%%%%%%%%%%%%%%%%%%%%%%%%%%%%%%%%%%%%%%%%%
\section{Conclusions}
The photonic sources required to build up optical quantum networks need to fulfill specific requirements for example with respect to their wavelength or bandwidth. These specifications are laid down in order to guarantee the compatibility with other hardware components, like the quantum memories, or to counteract undesired effects, like the dispersion in fibers. We investigate the properties of the PDC emission from a FP PPLN WG resonator having a \SI{2}{GHz} bandwidth at the telecommunication wavelengths. Our photonic source is working in the pulsed regime but on the edge towards narrower bandwidths. We performed a simulation of the joint spectral properties of our FP PPLN WG resonator and carried out a 2D Fourier transformation into the time domain to understand the temporal characteristics of the generated PDC emission. After experimentally demonstrating that our source produces photons in pairs in the SRM case we measured the time bins of signal and idler with and without spectral filtering that selects an individual resonator mode. The fringes in the time bins in the MRM case are connected to the FSR of the used resonator. Moreover, an exponential decay in a close to one nanosecond scale is observed in the time bins of signal and idler. With spectral filtering the fringe pattern disappears indicating that PDC is emitted into an individual resonator mode only. We observe excellent agreement between the experiment and theory, which correctly predicts the properties of the time bins. The different time-bin durations are comprehended by the different optical losses for the cross-polarized signal and idler, which in our case have an order of magnitude difference. We believe that the temporal properties of photon pairs play a crucial role in future quantum optical networks that rely on quantum interference between indistinguishable and independent light sources. Our results show the necessity of the temporal engineering of the quantum photonic sources.

\begin{backmatter}
%\bmsection{Funding}
\bmsection{Acknowledgment}
We thank Felix Schwenk for the help in determining the timing jitter of our single-photon detection system.
\bmsection{Disclosures}
The authors declare no conflicts of interest.
\bmsection{Data availability} Data underlying the results presented in this paper are not publicly available at this time but may be obtained from the authors upon reasonable request.
%\bmsection{Supplemental document}
%A supplemental document must be called out in the back matter so that a link can be included. For example, “See Supplement 1 for supporting content.” Note that the Supplemental Document must also have a callout in the body of the paper.
\end{backmatter}

%%%%%%%%%%%%%%%%%%%%%%% References %%%%%%%%%%%%%%%%%%%%%%%%%
%%%%%%%%%% If using BibTeX:
\bibliography{sample2}

\end{document}